\documentstyle[12pt]{article}

\begin{document} 

\begin{flushright}
KGKU-99-01 / MIE-99-01 \\
hep-ph/9902472 \\
Revised Version 
\end{flushright}

\vspace{10mm}

\begin{center}
{\large \bf Self-Dual Point and Unification }

\vspace{20mm}

Yoshikazu ABE 
            \footnote{E-mail address: abechan@phen.mie-u.ac.jp}, 
Chuichiro HATTORI$^a$ 
            \footnote{E-mail address: hattori@ge.aitech.ac.jp}, 
Masato ITO$^b$ 
            \footnote{E-mail address: mito@eken.phys.nagoya-u.ac.jp}, \\
Mamoru MATSUNAGA 
            \footnote{E-mail address: matsuna@phen.mie-u.ac.jp}
and Takeo MATSUOKA$^c$ 
            \footnote{E-mail address: matsuoka@kogakkan-u.ac.jp}

\end{center}

\begin{center}
{\it 
Department of Physics Engineering, Mie University, Tsu, JAPAN 514-8507 \\
{}$^a$Science Division, General Education, Aichi Institute of 
Technology, Toyota, JAPAN 470-0392 \\
{}$^b$Department of Physics, Nagoya University, Nagoya, 
JAPAN 464-8602 \\
{}$^c$Kogakkan University, Nabari, JAPAN 518-0498 
}
\end{center}

\vspace{15mm}

\begin{abstract}
We hypothesize that the vacuum is spontaneously shifted from 
the self-dual point in the moduli space, 
at which all dimensions are compact, 
and that the true vacuum results from the decompactification 
of the four space-time dimensions and remains in the vicinity 
of the self-dual point for extra dimensions. 
On the basis of our hypotheses we explore new types of the 
coupling unification which are not realized in the 
four-dimensional effective theory but in the framework of the 
higher dimensional theory. 
In the $SU(6) \times SU(2)_R$ string model, if the $SU(6)$ gauge group 
lives in the world-volume of 9-branes and the $SU(2)_R$ in 
the world-volume of 5-branes, or {\it vice versa}, 
unification solutions can be found in the vicinity of 
the self-dual point. 
\end{abstract}

\newpage 
\section{Introduction}
The recent advances in string theory have been stimulated by 
new concepts of non-perturbative string dualities 
and D-branes.\cite{Dbrane}-\cite{Fth} 
Using the string dualities it has been suggested that all 
five known perturbative superstring theories correspond 
to different particular limits of the underlying 
eleven-dimensional theory (M-theory), or 
twelve-dimensional theory (F-theory). 
In addition, in string theory there exist higher dimensional 
extended solitonic objects (D-branes). 
In the light of the dualities it is known that there is no absolute 
distinction between strings and solitonic objects. 
Therefore, the underlying theory is not simply a theory of string 
but a theory of extended objects including both strings and branes.

New aspects of string phenomenology have opened up due to 
recent developments in string theory. 
In fact, it has been found that there are new regions of the moduli space 
which are not contained in perturbative string theories. 
For instance, it is possible that some part of the gauge group and 
the matter content in the low-energy effective theory has a non-perturbative 
origin and that the other part of them has a perturbative origin. 
In addition, the string scale could lie somewhere between $\sim 1$TeV 
and the Planck scale $M_{\rm Pl}$. 
It should be emphasized that the coupling unification is properly realized 
in the framework of the higher-dimensional underlying theory. 
This does not necessarily mean that gauge couplings in the 
four-dimensional effective theory join at a certain energy scale. 
As a consequence, various paths seem to be possible for connecting 
the higher-dimensional underlying theory with low energy physics. 
Among them, the scenario of large extra dimensions has been 
extensively studied.\cite{ladim1}-\cite{ladim8} 
However, since proton stability suggests the existence of 
a large energy scale, 
it is unlikely that all extra dimensions are large in radius. 
The coexistence of large and small extra dimensions results in 
the fine-tuning problem.

The purpose of this paper is to explore new types of unification 
solutions without the fine-tuning that provide an alternative to 
the scenario of large extra dimensions. 
From the viewpoint of the unified theory, 
it is likely that the universe starts from a special point in 
the moduli space at which the symmetry of the underlying 
theory is maximally enhanced. 
This point should be a self-dual point in the moduli space 
for all dimensions, including the four space-time dimensions. 
It is hypothesized that the maximal symmetry is spontaneously 
broken due to some non-perturbative dynamics and that 
the decompactification occurs only for the four space-time dimensions. 
The beginning of the decompactification corresponds to the Big Bang. 
On the other hand, it is presumed that the non-perturbative 
dynamics also affects the sizes of the other dimensions, but 
the rate of their change is of order unity. 
Namely, as for extra dimensions other than the four space-time 
dimensions, the true vacuum remains in the vicinity of 
the self-dual point. 
Therefore, as far as extra dimensions are concerned, 
we are free from the fine-tuning problem. 
In this paper, based on our hypotheses we explore new types 
of the coupling unification that are not realized in the 
four-dimensional effective theory but only in the framework of 
the higher-dimensional theory.

This paper is organized as follows. 
In section 2 we give some hypotheses concerning the symmetry of 
the underlying theory. 
As mentioned above, we expect that starting from the self-dual 
point in the moduli space for all dimensions, the vacuum is 
shifted from that point. 
However, the true vacuum resides in the vicinity of the self-dual 
point for extra dimensions. 
In section 3 the unification problem is discussed 
in the framework of the Type IIB orientifold. 
On the basis of the brane picture, we explore new types of 
the coupling unification that are not contained in 
perturbative string theories. 
In the $SU(6) \times SU(2)_R$ string model, if the $SU(6)$ gauge group 
lives in the world-volume of 9-branes and the $SU(2)_R$ in 
the world-volume of 5-branes, or {\it vice versa}, 
we find unification solutions in the vicinity of 
the self-dual point in the moduli space. 
The final section is devoted to summary and discussion.

\section{Self-dual point and decompactification}
In this section we give hypotheses with regard to the underlying theory. 
If we follow the viewpoint of the unified theory thoroughly, 
it is natural that the underlying theory has the maximal 
symmetry allowed in its own theoretical framework. 
This implies that the vacuum of the underlying theory resides 
at the self-dual point in the moduli space for all dimensions, 
including the four space-time dimensions. 
This hypothesis is reasonable in view of the fact that in the 
heterotic superstring theory the maximal enhancement of the gauge 
symmetry occurs when the 16-dimensional compact space is an even 
self-dual lattice, 
which is required by the modular invariance. 
The hypothesis that the ground state of string theory lies at a 
point of maximally enhanced symmetry has been discussed 
by Dine et al.
\cite{Dine} 
They have assumed that this point corresponds to the self-dual 
point with respect to the modular transformation of moduli fields 
in the framework of the four-dimensional effective theory.
On the other hand, according to our hypothesis the self-dual point is 
concerned with the $S$- and $T$-duality transformations defined 
in the framework of the higher-dimensional underlying theory. 
Furthermore, as is discussed below, the emphasis is placed on 
the D-brane configurations in the vacuum.

In the $S$-duality, the string coupling constant $g_{\rm st}$ is 
transformed as 
\begin{equation}
     g_{\rm st} \longrightarrow g_{\rm st}' = \frac{1}{g_{\rm st}},
\end{equation}
where $g_{\rm st} = \exp (\langle \phi \rangle)$ and $\phi$ is 
the dilaton field. 
Thus the self-dual point with respect to the $S$-duality is 
\begin{equation}
     g_{\rm st} = 1, 
\end{equation}
which implies $\langle \phi \rangle = 0$. 
In the framework of F-theory, the $S$-duality transformation is 
contained in the modular transformation for the eleventh and the 
twelfth coordinates.\cite{Fth} 
On the other hand, under the $T$-duality transformation, 
the size $R$ of the compact space and the string coupling constant are 
transformed as 
\begin{eqnarray}
      R   & \longrightarrow & \widetilde{R} = \frac{\alpha '}{R}, \\
   g_{\rm st} & \longrightarrow & \widetilde{g}_{\rm st} = 
                            g_{\rm st} \frac{\sqrt{\alpha '}}{R}. 
\end{eqnarray}
In what follows we use the notation $M_*^{-2}$ for $\alpha '$. 
The mass scale $M_*$ represents the fundamental scale of the theory. 
With this notation the self-dual point with respect to 
the $T$-duality is given by 
\begin{equation}
    R = M_*^{-1}.
\end{equation}
At this point we have $\widetilde{g}_{\rm st} = g_{\rm st}$.

The first ansatz is that the universe started from the above 
self-dual point with respect to both the $S$-duality and 
the $T$-duality, at which all dimensions are compact and 
their size and shape are appropriately quantized. 
The second ansatz is that some portion of the maximal symmetry 
was spontaneously broken due to some unknown dynamics 
and that the universe was shifted from the self-dual point. 
Depending upon the shifted location of the vacuum in 
the moduli space, 
we have various types of the higher-dimensional theory. 
The five known ten-dimensional superstring theories correspond to 
different types of the decompactification of ten dimensions. 
The true vacuum results from the decompactification of 
only the four space-time dimensions; 
that is, in the true vacuum, only four dimensions began expanding, 
and their sizes are approximately $1.2 \times 10^{10}$\,light-yr. 
As for the other dimensions it is assumed that the universe 
remains in the vicinity of the self-dual point. 
In the context of the non-perturbative dynamics one should not ask 
why extra dimensions are compactified, but instead why the four space-time 
dimensions are decompactified.

The situation after the Big Bang is described by the 
four-dimensional effective theory. 
The size and shape of the compact space can be expressed in terms 
of the moduli fields $T_i$ in the effective theory. 
In fact, the vacuum expectation values (VEVs) of these fields 
are related to the sizes of the compact space along 
different directions. 
The moduli fields $T_i$ carry the charges of the family symmetry. 
As is well known, in nature the family symmetry is broken. 
For instance, since Yukawa couplings have hierarchical family 
structure, 
it is obvious that the family symmetry is broken in Yukawa 
couplings. 
Therefore, the moduli fields $T_i$, 
which have different quantum numbers of the family symmetry, 
do not necessarily have the same VEV. 
This means that the universe now be shifted from the self-dual 
point with respect to the $T$-duality for extra dimensions 
and that the sizes of the compact space may differ, 
depending on the directions. 
The second ansatz mentioned above is that the universe 
resides in the region of the moduli space as 
\begin{equation}
    g_{\rm st} = O(1) \quad {\rm and} \quad RM_* = O(1). 
\label{eqn:VSDP}
\end{equation}
In the next section, by assuming $g_{\rm st} \sim 1$, we explore 
unification solutions consistent with Eq.~(\ref{eqn:VSDP}).

Based on the above ansatze we explore new types of unification 
in which different types of D-branes possibly coexist. 
On the other hand, in Ref.\cite{Dine} the coexistence of different 
types of D-branes has not been taken into account. 
Hence the discussion in Ref.\cite{Dine} does not mention 
the possibility that the standard model is embedded separately 
into different types of D-branes.

\section{New types of the coupling unification}
The unification of gauge couplings and the gravitational coupling is 
properly realized in the framework of the higher-dimensional theory. 
The scales which appear in the four-dimensional effective theory 
do not necessarily represent the fundamental scale by itself. 
This implies that there possibly exist new types of the coupling unification 
that are realized in the framework of the higher-dimensional  theory
but not in the four-dimensional effective theory. 
To be specific, we take the Type IIB orientifold formulation, which 
corresponds to the Type I string theory. 
The bosonic part of the ten-dimensional action for the 
Type IIB orientifold\cite{orienti6} \cite{orienti4} 
\cite{orienti4D3} is 
\begin{equation}
    S_{10} = - \int \frac{d^{10}x}{(2\pi)^7} \sqrt{-g} \left(
         \frac{M_*^8}{g_{\rm st}^2}\, R + \frac{M_*^6}{g_{\rm st}} \,
            \frac{1}{4} F_{(9)}^2 + \cdots  \right),
\end{equation}
where $F_{(9)}$ refers to the gauge field coming from the 32 
9-branes. 
Generally, when the six-dimensional compact space is moded out by 
the discrete subgroup of $SO(6)$, 
the anomaly of R-R charges emerges. 
The consistency of the theory requires the cancellation of 
the anomaly, and then D-branes with various configurations 
should be added. 
This type of string vacuum is expected to be dual to 
a non-perturbative heterotic vacuum. 
In this case the gravity lives in the ten-dimensional bulk. 
On the other hand, the gauge groups live in the world volume 
of D$p$-branes with $p=3, 5, 7$ and $9$. 
Several authors\cite{orienti4}\cite{Aldaza} have constructed 
four-dimensional $Z_N$ and $Z_N \times Z_M$ orientifolds with 
different configurations of 5-branes, 7-branes and 9-branes. 
D-brane configurations are contrained strongly, 
depending on the orientifold group. 
If both D$p$-branes and D$q$-branes appear in the theory, 
the condition 
\begin{equation}
    p - q \equiv 0, \qquad  {\rm mod} \ 4 
\end{equation}
should be satisfied in order to preserve $N=1$ SUSY in the 
four-dimensional effective theory.\cite{Dbrane} 
In this paper we do not discuss the dependence of the D-brane configuration 
on the orientifold group. 
Instead, we study unification solutions from the phenomenological 
point of view on the supposition of some appropriate D-brane configurations. 
After dimensional reduction down to four dimensions, 
we obtain the bosonic part of the action, 
\begin{equation}
    S_4 = - \int \frac{d^{4}x}{2\pi} \sqrt{-g} \left(
         \frac{M_*^8 \,V_6}{g_{\rm st}^2 \,(2\pi)^6} \,R + 
         \sum _{p=3}^9 \frac{M_*^{p-3} \,V_{p-3}}{g_{\rm st} \,(2\pi)^{p-3}} 
           \, \frac{1}{4} F_{(p)}^2 + \cdots  \right),
\label{eqn:S4}
\end{equation}
where the second terms represent the contribution of different 
D$p$-branes with $p=3, 5, 7$ and $9$ and 
$V_{p-3}$ is the $(p-3)$-dimensional volume occupied by 
the D$p$-brane in the compact space.

The new concept of D-branes opened up diverse paths of connecting 
the higher-dimensional  underlying theory with low energy physics. 
Let us explore new types of the unification which are not 
contained in perturbative string theories. 
It is desirable for us to obtain unification solutions without 
fine-tuning. 
We consider two cases, one in which the standard model 
is embedded into a single type of branes and one in which it is not. 

\vskip 5mm

\begin{flushleft}
Case 1 
\end{flushleft}

In this case the standard model is embedded into a single type of 
branes and then four-dimensional gauge couplings meet 
at a certain scale ($M_{\rm GUT}$). 
Therefore, this type of gauge unification is similar to 
conventional unification. 
From Eq.~(\ref{eqn:S4}) we have the relations 
\begin{eqnarray}
    M_{\rm Pl}^2 & = & 8 g_{\rm st}^{-2} \frac{V_6}{(2\pi)^6} M_*^8, \\
    \alpha _G^{-1} & = & 2 g_{\rm st}^{-1} 
                      \frac{V_{p-3}}{(2\pi)^{p-3}} M_*^{p-3}. 
\end{eqnarray}
Since the vacuum considered here is on the border of 
the perturbative region of the moduli space, 
these relations may be subject to radiative corrections. 
However, since the assumed maximal symmetry of the underlying 
theory implies that massive spectra have $N=4$ SUSY structure, 
it is expected that the tree-level relations are applicable 
even in the non-perturbative region of 
the moduli space.\cite{ladim1}

For simplicity, for the moment we neglect the difference of 
the sizes along various directions in the compact space. 
In such a case, the $(p-3)$-dimensional volume can be simply 
expressed as 
\begin{equation}
     V_{p-3} = (2\pi R)^{p-3}. 
\end{equation}
By taking $g_{\rm st} \sim 1$ and $\alpha _G^{-1} \sim 24$  as inputs, 
we obtain 
\begin{eqnarray}
   & & (R M_*)^{p-3} \sim 12, \\
   & & R^{-1} \sim \frac{1}{2\sqrt{2}} (12)^{-4/(p-3)} M_{\rm Pl}. 
\label{eqn:Rinv}
\end{eqnarray}
Furthermore, when the perturbative unification of the gauge couplings 
is realized at the scale $M_{\rm GUT} \sim 2 \times 10^{16}$ GeV 
as in the MSSM, 
it is natural for $R^{-1}$ to be identified as $M_{\rm GUT}$. 
If this is the case, Eq.~(\ref{eqn:Rinv}) holds only for 
\begin{equation}
   p = 5, 
\end{equation}
and we have 
\begin{equation}
   R M_* \sim 3.5, \qquad M_* \sim 7 \times 10^{16}{\rm GeV}. 
\end{equation}
This solution implies that the standard model gauge group 
lives in the world-volume of 5-branes and that the size of 
the compact space is $O(1)$ in $M_*^{-1}$ units. 
The Planck scale $M_{\rm Pl}$ is larger than $M_*$ by about two 
orders of magnitude.  
This is attributable to the difference between the dimensions 
of the branes. 
Recently, proton stability has been restudied in SUSY-GUT models.
\cite{Goto} 
In these studies it was found that colored Higgs masses should be larger than 
$M_{\rm GUT} \sim 2 \times 10^{16}$ GeV. 
This suggests that the perturbative unification of gauge couplings 
at $M_{\rm GUT} \sim 2 \times 10^{16}$ GeV, such as in the MSSM, 
may be accidental. 
On the supposition that the unification bears no resemblance to the 
conventional one, 
we proceed to study the second case.

\vskip 5mm

\begin{flushleft}
Case 2 
\end{flushleft}

In this case the standard model is not embedded into 
a single type of brane. 
More concretely, we consider the case in which two gauge 
groups are not unified in the four-dimensional effective theory. 
The gauge group $G$ at the unification scale is written as 
\begin{equation}
     G = G_p \times G_q. 
\end{equation}
The gauge group $G_p$ $(G_q)$ lives in the world-volume of the $p$ $(q)$-brane. 
Similarly to the previous case, we have the relations 
\begin{eqnarray}
    M_{\rm Pl}^2 & = & 8 g_{\rm st}^{-2} \frac{V_6}{(2\pi)^6} M_*^8, \\
    \alpha _p^{-1} & = & 2 g_{\rm st}^{-1} 
                      \frac{V_{p-3}}{(2\pi)^{p-3}} M_*^{p-3}, \\
    \alpha _q^{-1} & = & 2 g_{\rm st}^{-1} 
                      \frac{V_{q-3}}{(2\pi)^{q-3}} M_*^{q-3}. 
\end{eqnarray}
As mentioned above, we require the condition 
\begin{equation}
    p - q \equiv 0, \qquad  {\rm mod} \ 4. 
\end{equation}
If both $\alpha _p^{-1}$ and  $\alpha _q^{-1}$ are larger than 
$2 g_{\rm st}^{-1}$, 
then $p, q \neq 3$. 
Therefore, if we assume $p \neq q$, the solutions become 
\begin{equation}
   (p,\ q) = (9,\ 5), \quad  (5,\ 9). 
\end{equation}
Note that the 9-brane and 5-brane can be interchanged by a $T$-duality 
transformation with respect to four dimensions in the six-dimensional 
compact space. 
Thus, if the vacuum is self-dual with respect to this $T$-duality, 
then $G_p = G_q$ should be satisfied. 
Conversely, if $G_p \neq G_q$, the vacuum is not self-dual with 
respect to the $T$-duality. 
If $p = q$, we have $p = q = 5, 7$. 
Although the dimensions of the two branes are the same in these cases, 
the brane configurations should be different. 
In this paper we do not discuss these cases.

Here we take up a phenomenologically viable string model with 
the gauge group $SU(6) \times SU(2)_R$, 
which can explain the hierarchical pattern 
of quark-lepton masses and mixings systematically.\cite{ours} 
The gauge groups $SU(3)_c$ and $SU(2)_L$ in the standard model are 
included in this $SU(6)$, but $U(1)_Y$ is not. 
In spite of such attractive results of this model, 
we had not been successful in obtaining the perturbative unification 
of gauge couplings. 
As a matter of fact, the analysis in Ref.\cite{ours} shows that 
the numerical values of the gauge couplings are 
\begin{equation}
     \alpha (SU(6))^{-1} \sim 16, \qquad 
     \alpha (SU(2)_R)^{-1} \sim 10 
\end{equation}
at the scale $(0.5 \sim 1) \times 10^{18}$ GeV. 
Let us apply these results to the present framework with 
$g_{\rm st} \sim 1$. 
By taking $G_p = SU(6)$, $G_q =SU(2)_R$ and $(p,\ q)=(9,\ 5)$, 
we obtain 
\begin{eqnarray}
    M_{\rm Pl}^2 & \sim &  8 \frac{V_6}{(2\pi)^6} M_*^8, \\
        16   & \sim &  2 \frac{V_6}{(2\pi)^6} M_*^6, \\
        10   & \sim &  2 \frac{V_2}{(2\pi)^2} M_*^2. 
\end{eqnarray}
The geometrical average of the size of the two extra dimensions 
on which the 5-brane lives is given by 
\begin{equation}
    R_2 = \frac{1}{2\pi} \sqrt {V_2} \sim 2.2 \times M_*^{-1}.
\end{equation}
The geometrically averaged size of the four extra dimensions 
perpendicular to the 5-brane becomes 
\begin{equation}
    R_4 = \frac{1}{2\pi} \left( \frac{V_6}{V_2} \right)^{1/4} 
             \sim 1.1 \times M_*^{-1}.
\end{equation}
In this case the fundamental scale is 
\begin{equation}
   M_* \sim 1.5 \times 10^{18} \ {\rm GeV}. 
\end{equation}
The value $R_2^{-1} \sim 0.7 \times 10^{18}$ GeV is consistent 
with the result in Ref.\cite{ours}. 
It should be noted that in the present solution the sizes of the 
six-dimensional compact space ($R_2$ and $R_4$) is $O(1)$ in $M_*^{-1}$ units. 
In the narrow energy region ranging from $R_2^{-1}$($R_4^{-1}$) to 
$M_*$ we have the contributions of KK-modes in the $R_2$($R_4$)-direction. 
In the energy region below $R_2^{-1}$, we can neglect 
the contributions of the KK-modes and the winding modes and obtain 
a four-dimensional effective theory. 
Furthermore, when $(p,\ q)=(5,\ 9)$, we obtain 
\begin{eqnarray}
     R_2 & \sim & 2.8 \times M_*^{-1}, \\
     R_4 & \sim & 0.9 \times M_*^{-1}, \\
     M_* & \sim & 1.9 \times 10^{18} \ {\rm GeV}. 
\end{eqnarray}
%

\section{Summary and discussion}
In this paper we have made some hypotheses regarding the higher-dimensional  
underlying theory. 
The first is that the underlying theory has the maximal symmetry 
allowed in its own theoretical framework. 
This implies that the universe starts from the self-dual point 
with respect to both the $S$-duality and the $T$-duality, 
at which all dimensions are compact. 
The second is that, due to some dynamics, the decompactification 
occurs for the four space-time dimensions and that the true vacuum 
remains in the vicinity of the self-dual point for extra 
dimensions. 
The unification of gauge couplings and the gravitational coupling 
is realized properly in the framework of the higher-dimensional 
theory with strings and D-branes. 
We explored new types of unification solutions which are not 
contained in perturbative string theories. 
If the standard model is embedded in a single type of D-brane, 
four-dimensional gauge couplings meet at a certain scale. 
On the other hand, if the standard model is not embedded in 
a single type of brane, 
we have new solutions of the coupling unification in which 
the four-dimensional gauge coulpings do not join at a certain scale. 
Based on our hypotheses we studied phenomenologically viable 
solutions of the unification with 
\begin{equation}
    g_{\rm st} \sim 1 \quad {\rm and} \quad RM_* = O(1). 
\end{equation}
It should be emphasized that in solutions of this type, 
as far as extra dimensions are concerned, 
we are free from the fine-tuning problem. 
In the $SU(6) \times SU(2)_R$ string model, if the $SU(6)$ gauge group 
lives in the world-volume of 9-branes and the $SU(2)_R$ in 
the world-volume of 5-branes, or {\it vice versa}, 
we find such unification solutions in the vicinity of 
the self-dual point in the moduli space. 
In this paper we did not discuss the dependence of the D-brane 
configuration on the orientifold group. 
A study of this subject will be made elsewhere.

In this study we concentrated on the gauge group and gauge 
couplings. 
It is also important to determine what types of charged matter 
chiral superfields appear in the four-dimensional effective 
theory, depending on the brane configuration in the vacuum. 
In the $SU(6) \times SU(2)_R$ string model, in which 9-branes 
coexist with 5-branes, 
we have three types of charged matter superfields. 
In the Type I formulation, they are 
(i) open strings starting and ending on 9-branes, 
(ii) open strings starting on 9-branes and ending on 5-branes, 
and (iii) open strings starting and ending on 5-branes. 
When the gauge groups $G_9$ and $G_5$ on 9-branes and 5-branes 
correspond to the $SU(6)$ and $SU(2)_R$, respectively, 
the type (ii) strings transform as the bifundamental 
representation of the $SU(6) \times SU(2)_R$, 
which is denoted ${\bf (6^*, 2)}$. 
On the other hand, the type (i) ((iii)) strings 
are singlet under the $SU(2)_R$ ($SU(6)$) but are expressed as 
second rank tensors under the $SU(6)$ ($SU(2)_R$). 
In the Type I formulation it can be shown that these second rank 
tensors are antisymmetric. 
This results from the massless conditions, which can be expressed in terms of 
the inner products between the root vectors and shift vectors. 
Thus, the strings starting and ending on 9-branes (5-branes) 
transform as ${\bf(15, 1)}$ (${\bf (1, 1)}$) under 
the $SU(6) \times SU(2)_R$. 
Therefore, the charged matter chiral superfields become ${\bf (15, 1)}$ 
and ${\bf (6^*, 2)}$, which compose a fundamental representation 
{\bf 27} of $E_6$. 
Further, the $SU(6)$ gauge anomaly is cancelled within one set 
of the open strings. 
The present model has the same charged matter content as the 
perturbative heterotic string model. 
This result implies that the phenomenological analyses in Ref.\cite{ours} 
are applicable in the non-perturbative region of the moduli space.

At present it seems that there are a large number of possible paths 
for connecting the underlying theory with the standard model. 
In order to explore a realistic scenario, 
we need to solve many important problems on the basis of 
the brane picture. 
New perspectives in string phenomenology will open 
from further developments in the underlying theory.

\section*{Acknowledgements}
The authors would like to thank Dr. Y. Imamura 
for valuable discussions. 
One of the authors (T. M.) is supported in part by 
a Grant-in-Aid for Scientific Research, 
Ministry of Education, Science, Sports and Culture, Japan 
(Nos. 10140209 and 10640256).



\end{document}